\documentclass[twocolumn]{aastex6}
\usepackage{longtable}

\begin{document}
\slugcomment{Accepted for publication in ApJ}
\shortauthors{Lim et al.}
\shorttitle{Reversed radial distribution trend of subpopulations in the NGC~362 and NGC~6723}

\title{Reversed radial distribution trend of subpopulations in the globular clusters NGC~362 and NGC~6723}

\author{
Dongwook Lim\altaffilmark{1},
Young-Wook Lee\altaffilmark{1},
Mario Pasquato\altaffilmark{1},
Sang-Il Han\altaffilmark{2},
and
Dong-Goo Roh\altaffilmark{2}
}

\altaffiltext{1}{Center for Galaxy Evolution Research \& Department Astronomy, Yonsei University, Seoul 03722, Korea; dwlim@galaxy.yonsei.ac.kr, ywlee2@yonsei.ac.kr}
\altaffiltext{2}{Korea Astronomy and Space Science Institute, Daejeon 34055, Korea}

\begin{abstract}
Most globular clusters (GCs) are now known to host multiple stellar populations with different light element abundances. 
Here we use narrow-band photometry and low-resolution spectroscopy for NGC~362 and NGC~6723 to investigate their chemical properties and radial distributions of subpopulations. 
We confirm that NGC~362 and NGC~6723 are among the GCs with multiple populations showing bimodal CN distribution and CN-CH anti-correlation without a significant spread in calcium abundance.
These two GCs show more centrally concentrated CN-weak earlier generation stars compared to the later generation CN-strong stars.
These trends are reversed with respect to those found in previous studies for many other GCs.
Our findings, therefore, seem contradictory to the current scenario for the formation of multiple stellar populations, but mass segregation acting on the two subpopulations might be a possible solution to explain this reversed radial trend.
\end{abstract}

\section{Introduction}\label{intro}
Recent observations suggest that most globular clusters (GCs) host multiple stellar populations showing star-to-star abundance variations in the light elements, such as C, N, O, Na and Al \citep[e.g.,][and references therein]{Car09,Gra12,Pio15}.
Among several scenarios for the origin of these abundance variations, the most widely accepted one is the self-enrichment scenario, which explains these variations by the chemical pollution/enrichment from earlier generation stars, such as intermediate-mass asymptotic giant branch (IMAGB) stars \citep{VD08}, rotating AGB stars \citep{Dec09}, interacting binary stars \citep{de09}, and fast-rotating massive stars (FRMSs; \citealt{Dec07b}).
In this scenario, later generation stars are expected to be formed by the gas ejected from earlier generation stars in the innermost region of a proto GC \citep{Dec07a,D'er08,Ves13}.
Therefore, the later generation stars would be observed to be more centrally concentrated than earlier generation stars unless this radial distribution was seriously affected by dynamical evolution \citep[see, e.g.,][]{Mih15}.
This radial trend is indeed observed in many GCs, including  $\omega$~Cen, M13, NGC~3201, NGC~6752, and 47~Tuc \citep{Bel09,Kra10,Kra11,Lar11,Nat11,JP12,Mil12}.
The incidence of spectroscopic binaries in different subpopulations, which is less affected by the dynamical evolution, also supports that later generation stars were formed in a denser environment where binaries are efficiently destroyed, resulting in a lower binary fraction for later generation stars \citep{D'Or10,Luc15}.
\begin{figure*}
\centering
\includegraphics[width=0.9\textwidth]{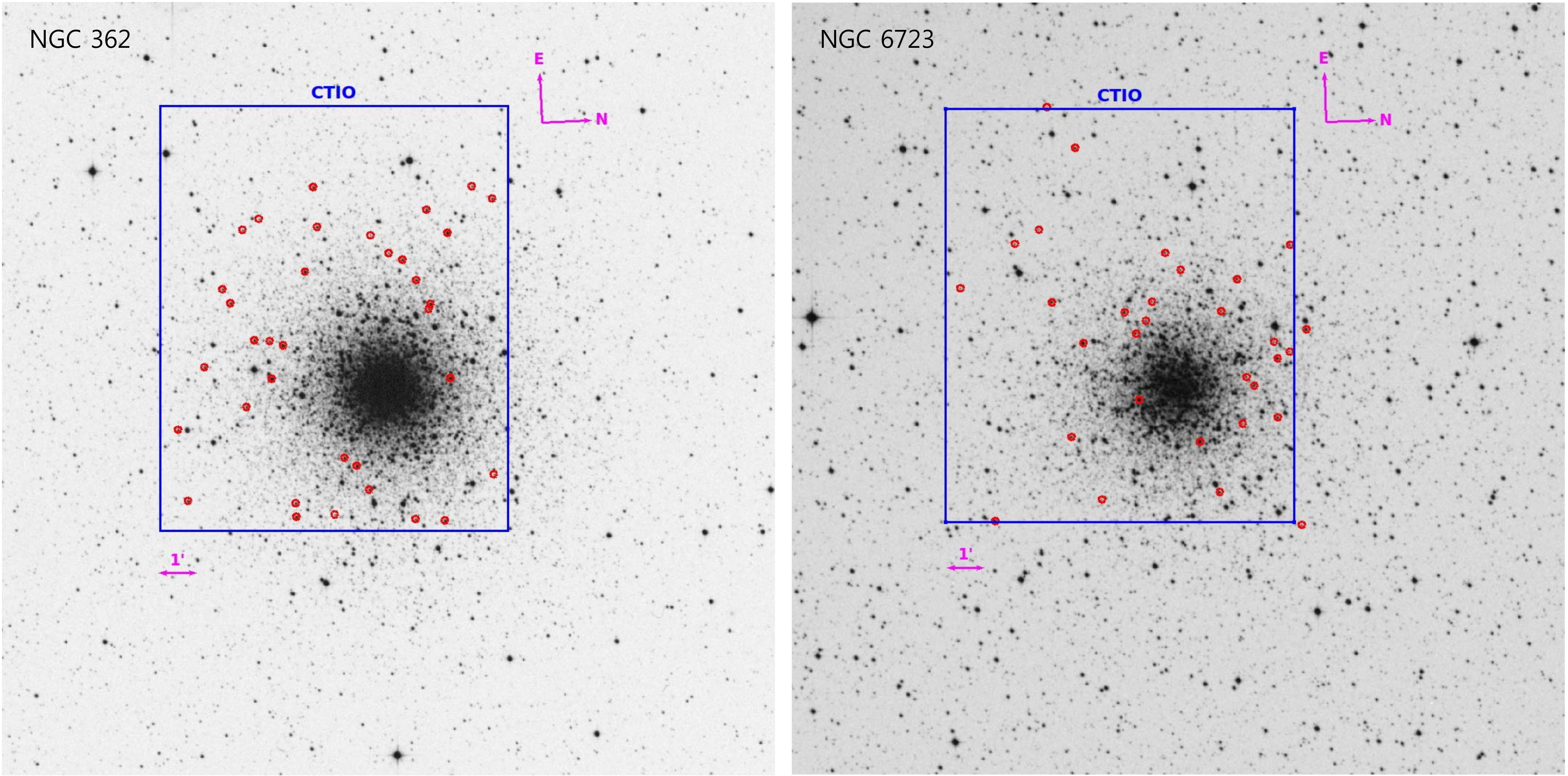}
\figcaption{Our observed fields (blue squares) on the STScI images of NGC~362 and NGC~6723.
The red circles indicate our spectroscopic target stars.
\label{fig_fov}
}
\end{figure*}

However, not every GC with multiple populations shows this general radial distribution trend. 
\citet{Car10} and \citet{jwlee15} report a central concentration of metal-poor earlier generation stars in NGC~1851\footnote{\citet{Mil09} showed that two stellar populations of sub-giant branch stars in NGC~1851 share the same radial distribution, which, however, is contradicted by \citet{Car11}.} and M22, respectively. 
These GCs are known as peculiar GCs showing intrinsic heavy elements dispersions \citep{jwlee09,Mar09,Car11,Lim15}, and therefore, their reversed radial trends may be considered to be the result of merging of two individual GCs \citep[see][]{Car11}.  
In the case of GCs without heavy elements spread, \citet{Dal14} have shown that two stellar populations in NGC~6362 share the same radial distribution, which is explained as a full spatial mixing accelerated by high mass loss rate of this GC (\citealt{Mih15}; see also \citealt{Muc16}).
Furthermore, \citet{Lar15} recently suggested that first generation stars (primordial group) are more centrally concentrated than second generation stars (enriched group) in M15 using $Hubble~Space~Telescope~(HST)$ WFC3 photometry.
This was not detected in the previous study for the same GC by \citet{Lar11} using the Sloan Digital Sky Survey (SDSS) data, which cover only the outer region of a cluster. 
Contrary to the cases of M22 and NGC~1851, the reversed radial trend in M15 is unlikely to be the result of merging, because this GC does not show Fe spread although variations in the light elements and some neutron-capture elements (Ba, Eu) were reported (\citealt{Sne97}; see also \citealt{BT16}).
The spatial mixing due to dynamical evolution is also unlikely to explain this reversed radial trend \citep[see, e.g.,][]{Ves13}.
Therefore, the presence of a reversed radial distribution trend in M15 casts some doubt on the current self-enrichment scenario, and thus a search for further instances of this radial characteristic is required.

Investigating the radial distribution of multiple stellar populations, however, is not a simple task because spectroscopic observations are hard to secure a large enough number of samples and it is difficult to divide subpopulations using photometric observation alone.
In this regard, our narrow-band photometry, combined with low-resolution spectroscopy, would be a useful tool for this investigation.
Our previous studies have shown that narrow-band photometry using ``Ca'' and ``Ca+CN'' filters can efficiently detect multiple stellar populations with different chemical properties, and this is confirmed by low-resolution spectroscopy \citep{Lim15,Han15}. 
In this study, we have investigated the chemical properties and radial distributions of stars in NGC~362 and NGC~6723 by employing the same techniques and report that these two GCs show a central concentration of earlier generation stars, similarly to the case of M15. 

\begin{figure*}
\centering
\includegraphics[width=1.0\textwidth]{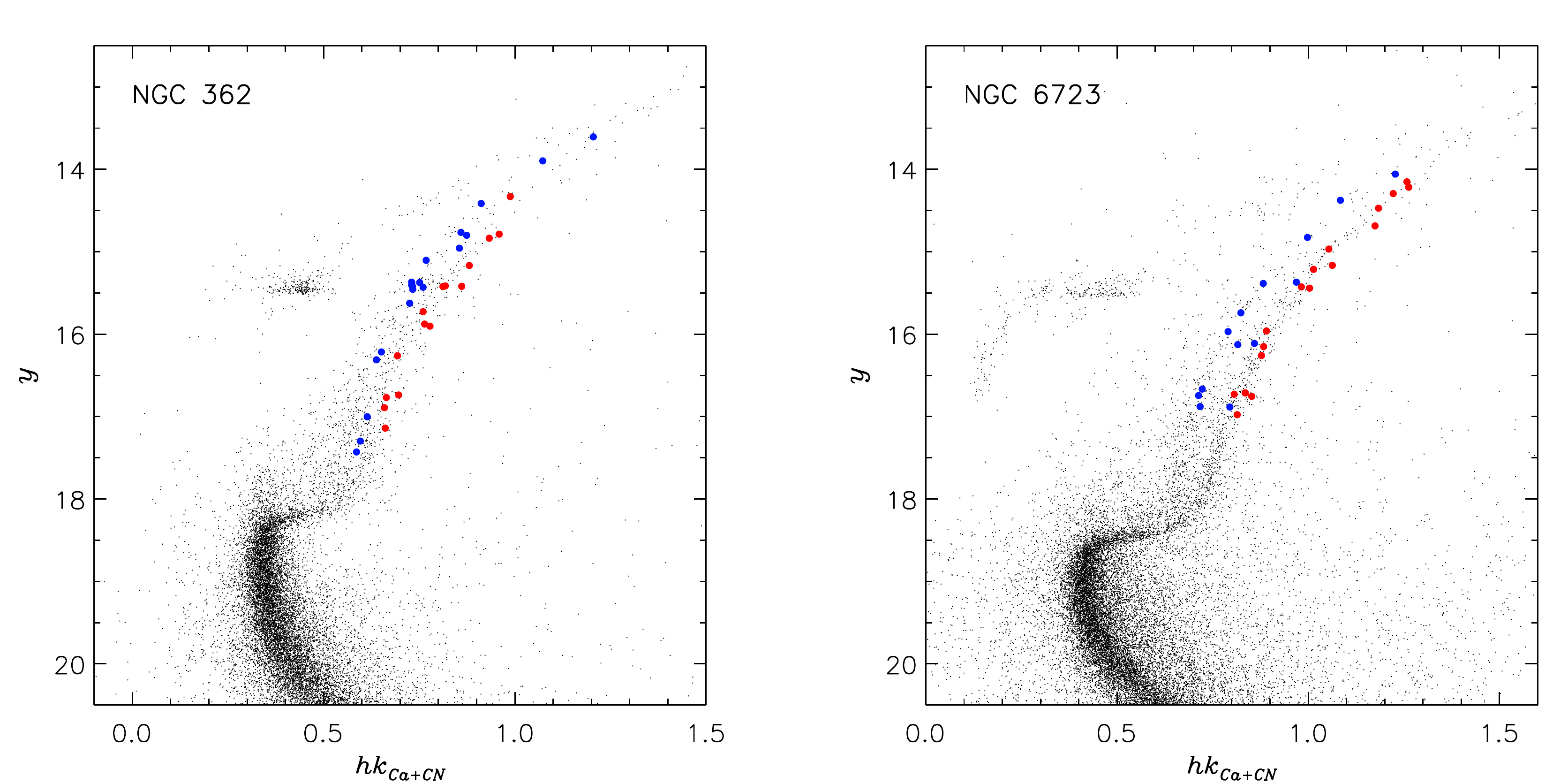}
\figcaption{
CMDs for NGC~362 (left) and NGC~6723 (right) in ($y$, $hk_{Ca+CN}$) plane obtained with the Ca+CN filter set at CTIO.
Spectroscopic target stars are also identified in these CMDs, where the blue and red circles are CN-weak and red CN-strong stars, respectively (see Section~\ref{spec}).
Note that RGB spread and split are shown in NGC~362 and NGC~6723, respectively.
\label{fig_cmd}
}
\end{figure*}
\begin{deluxetable}{cccccc}
\tabletypesize{\normalsize}
\tablewidth{0pt}
\tablecaption{Mask Descriptions and Spectroscopic Observation Log\label{tab_log}}
\tablehead{
\colhead{Object} & \colhead{Mask} & \colhead{No of stars} & \colhead{Exposures (N$\times$s)}
}
\startdata
NGC 362  & Bright    & 23 & 2$\times$1200 \\
         &           &    & 2$\times$1500 \\
         & Faint     & 26 & 4$\times$1500 \\
NGC 6723 & Bright    & 19 & 4$\times$1200 \\
         &           &    & 3$\times$1500 \\  
         & Faint     & 31 & 5$\times$1500 
\enddata
\end{deluxetable}
\section{Narrow-Band Photometry and Low-Resolution spectroscopy}\label{obs}
Our photometry was obtained at the Cerro Tololo Inter-American Observatory (CTIO) 4m Blanco telescope with ``Ca+CN'' filter in July 2009.  
As described in \citet{Lim15}, this filter was originally designed to measure only the strength of Ca II H\&K lines, however, due to the deterioration, the filter passband was shifted to include CN molecular band at 3883 {\AA}. 
Fortunately, this filter system became sensitive enough to detect the difference of CN band strength \citep[see][]{Hsyu14,Lim15}, and this could therefore be effectively used to study multiple stellar populations with different CN abundances.  

In this observation, we have used the MOSAIC II CCD Imager, which provides a pixel scale of 0.27{\arcsec} and a field of view (FOV) of 36{\arcmin} $\times$ 36{\arcmin}. 
However, only stars placed on chip 6 (FOV $\sim$ 9{\arcmin} $\times$ 18{\arcmin}) are used for the analysis to avoid possible chip-to-chip variations of the mosaic CCDs \citep[see][]{Han09,Roh11}.
The observed fields are shown in Figure~\ref{fig_fov}.
Similarly to our previous works \citep{Roh11,Han15,Lim15}, IRAF\footnote{IRAF is distributed by the National Optical Astronomy Observatory, which is operated by the Association of Universities for Research in Astronomy (AURA) under a cooperative agreement with the National Science Foundation.} MSCRED package and DAOPHOT II/ALLFRAME \citep{Ste87,Ste94} were used for preprocessing and point spread function photometry.   
We also used CHI and SHARP parameters to exclude non-stellar objects and bad samples (i.e., CHI $>$ 3.0, and SHARP $>$ $|1.0|$).
Note that, unlike our previous studies, the $sep$ index was not used for sample selection in order to include stars in the central region of a cluster. 
Finally, astrometry was performed using the IRAF FINDER package with the 2MASS All-Sky Point Source catalog.
Figure~\ref{fig_cmd} shows color-magnitude diagrams (CMDs) for NGC~362 and NGC~6723 in ($y$, $hk_{Ca+CN}$\footnote{We have used the same definition of the $hk$ index defined by \citet{Ant91}, $hk=($Ca$-b)-(b-y)$. However, in this study, this index is expressed as $hk_{Ca+CN}$ to clarify the difference of filter response function.}) plane.
In these CMDs, NGC~362 shows a spread on the red giant-branch (RGB), and NGC~6723 shows two distinct RGBs, suggesting variations in CN band strength among RGB stars. 
\begin{figure*}
\centering
\includegraphics[width=0.78\textwidth]{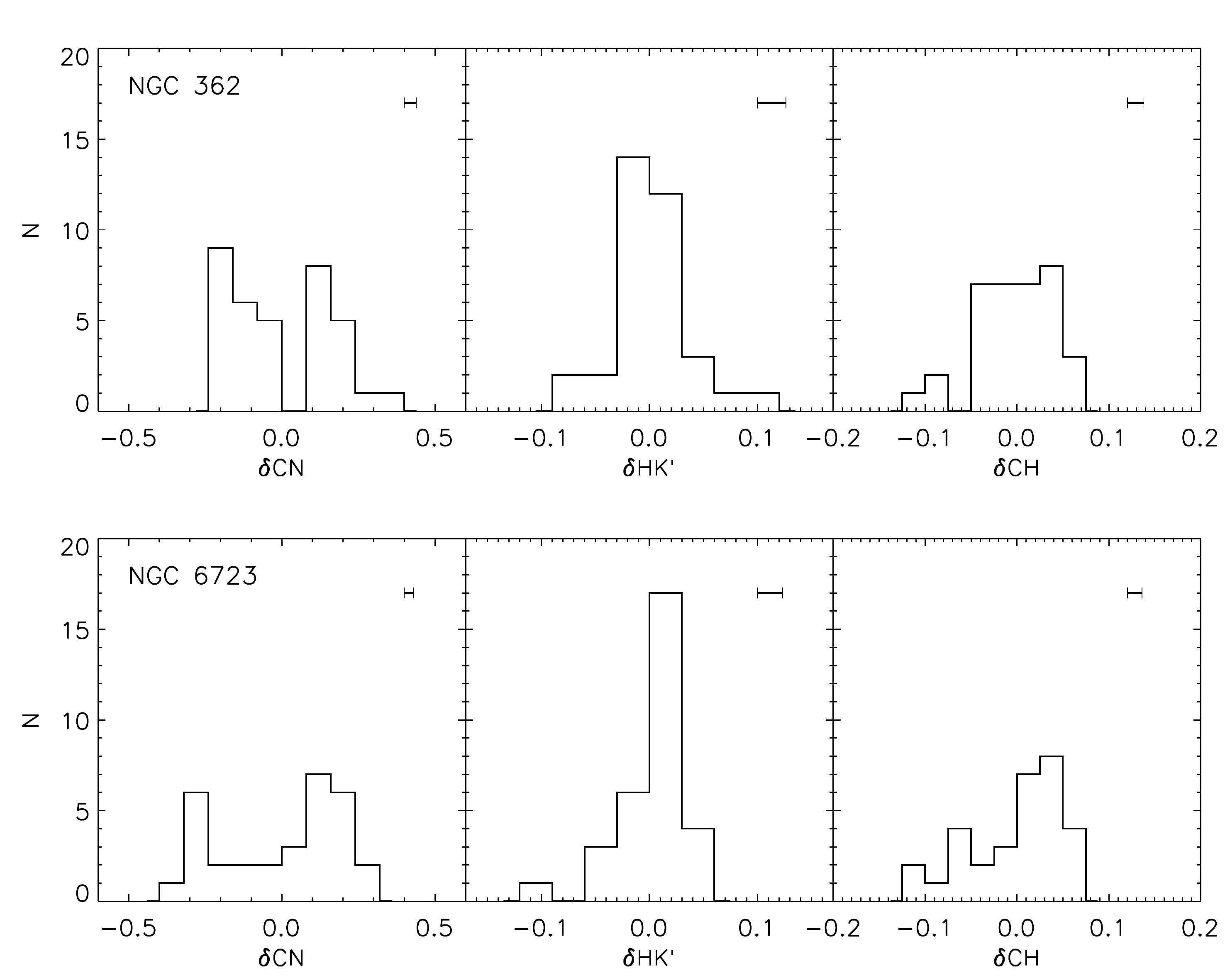}
\figcaption{
The histograms of $\delta$CN, $\delta$HK$'$, and $\delta$CH indices for RGB stars in NGC~362 (upper panels) and NGC~6723 (lower panels).
The horizontal bar denotes the typical measurement error (1$\sigma$).
Note that both GCs show the bimodal distribution in the $\delta$CN histogram.
\label{fig_hist}
}
\end{figure*}

The spectroscopic data were obtained from Las Campanas Observatory (LCO) 2.5m duPont telescope in July 2011 and June 2014. 
We used Wide Field Reimaging CCD Camera (WFCCD) with HK grism, providing a dispersion of 0.8 {\AA}/pixel and a central wavelength of 3700 {\AA}. 
Two multi-slit masks were made for each GC using the CTIO photometry data, and at least four exposures were taken for each mask (see Table~\ref{tab_log}).
For the data reduction, following \citet{Lim15}, the modified version of WFCCD reduction package and IRAF were used. 
More detailed information regarding observation and data reduction can be found in \citet{Lim15} and \citet{Pro06}. 
We also measured the radial velocity of each star using the $rvidlines$ task in IRAF RV package and estimated signal-to-noise (S/N) ratio at $\sim$3900 {\AA}.
From these parameters, non-member stars (radial velocity $>$ 2.5$\sigma$ of the mean velocity of each GC) and bad samples (S/N $<$ 8) were excluded from our analysis. 
Finally, we have obtained spectra for 35 stars in NGC~362, and 31 stars in NGC~6723, which are identified in Figures~\ref{fig_fov} and \ref{fig_cmd}.

\begin{figure}
\centering
\includegraphics[width=0.48\textwidth]{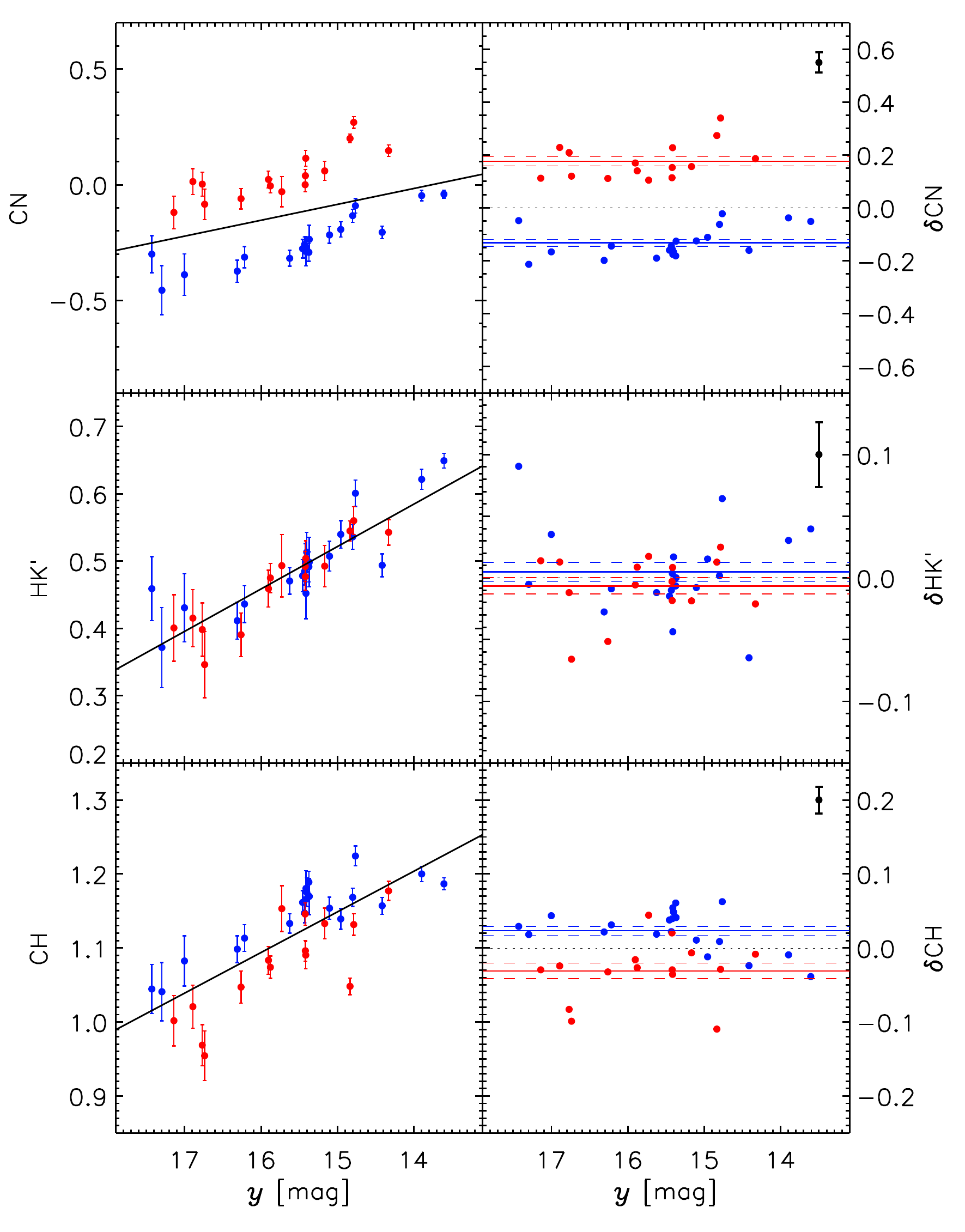}
\figcaption{
Left panels: Measured spectral indices (CN, HK$'$, and CH) as functions of $y$ magnitude for RGB stars in NGC~362, where the blue and red circles are CN-weak and CN-strong stars.
Right panels: The $\delta$CN, $\delta$HK$'$, and $\delta$CH indices plotted against $y$ magnitude.
The mean value and the error of the mean ($\pm$1$\sigma$) for each subpopulation are denoted by solid and dashed lines, respectively.
The vertical bars in the upper right corner denote the typical measurement error for each index.
Note that the two subpopulations are clearly separated in the $\delta$CN and $\delta$CH indices, but not in the $\delta$HK$'$ index.
In addition, the strengths of the CN and CH bands are anti-correlated. 
\label{fig_n362spec}
}
\end{figure}

\begin{figure}
\centering
\includegraphics[width=0.48\textwidth]{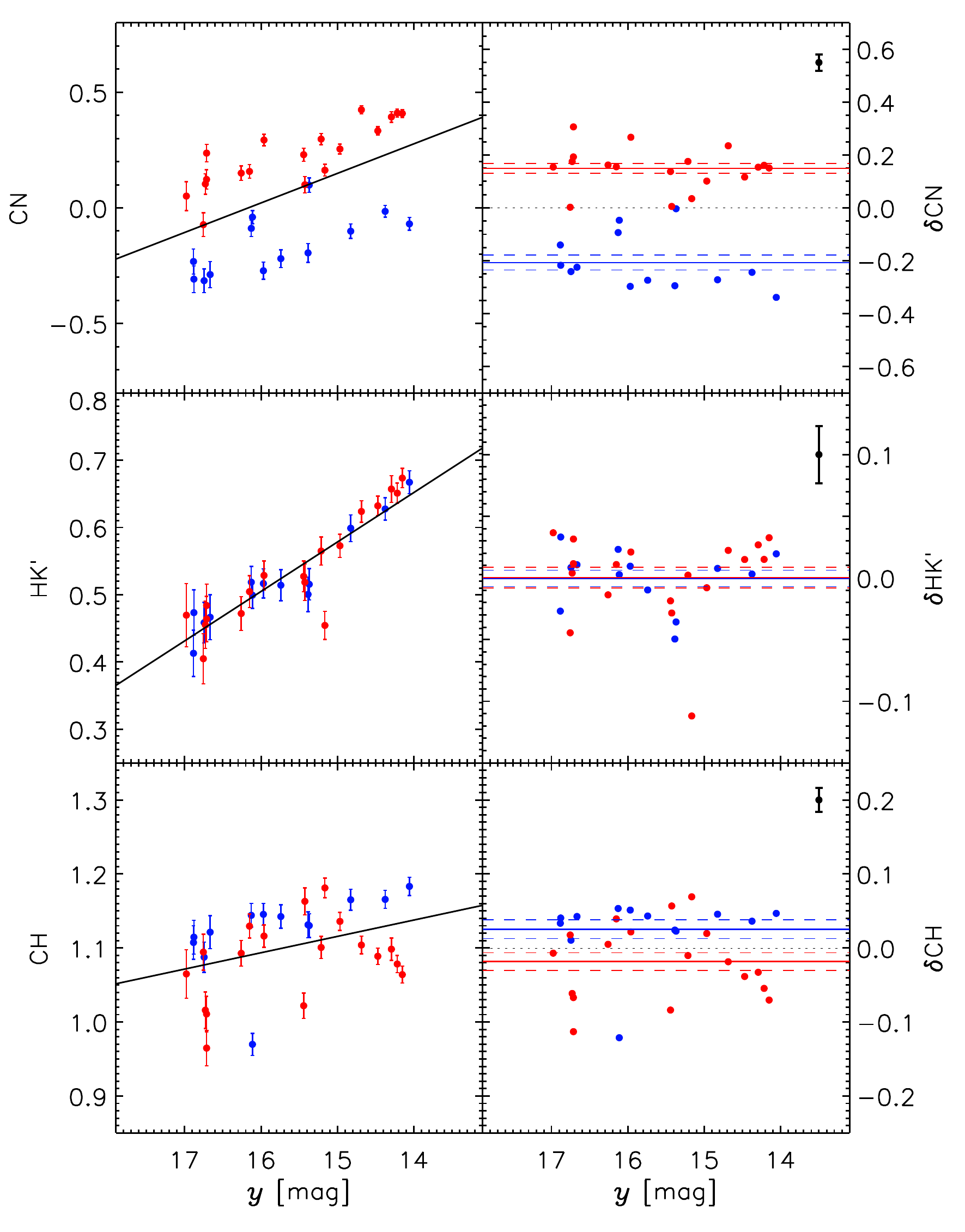}
\figcaption{
Same as Figure~\ref{fig_n362spec}, but for NGC~6723. 
Note that this GC shows similar trends as NGC~362, such as CN bimodality and CN-CH anti-correlation.
\label{fig_n6723spec}
}
\end{figure}

After the data reduction, we measured the CN, HK$'$, and CH indices for each star, which are defined by \citet{Har03} and \citet{Lim15}. 
The definitions for these indices are
\begin{eqnarray*} 
{\rm HK'}  & = & -2.5 \log{\frac{F_{3916-3985}}{2F_{3894-3911}+F_{3990-4025}}} , \\
{\rm CN}(3839) & = & -2.5 \log{\frac{F_{3861-3884}}{F_{3894-3910}}} , \\
{\rm CH4300} & = & -2.5 \log{\frac{F_{4285-4315}}{0.5F_{4240-4280}+0.5F_{4390-4460}}} ,
\end{eqnarray*}
where $F_{3916-3985}$, for example, is the integrated flux from 3916 to 3985 {\AA}.
In addition, delta indices ($\delta$CN, $\delta$HK$'$, and $\delta$CH) for each spectral index are also derived as the difference between the original index and the least squares fitting of the full sample in a GC (black solid lines in left panels of Figures~\ref{fig_n362spec} and ~\ref{fig_n6723spec}) to reduce the effects of effective temperature ($T_{\rm eff}$) and surface gravity ($\log g$). 
The measured spectral indices are listed in Table~\ref{tab_index}.

\section{Multiple Stellar Populations with Different CN strengths}\label{spec}
Since the early study by \citet{Smi84}, several studies have reported variations in the CN and CH molecular bands among stars in NGC~362 \citep{Kay08,SL09}. 
More recent studies found Na-O anti-correlation, suggesting multiple stellar populations (\citealt{Car13}; see also \citealt{SK00}), but no evidence for heavy elements spread was reported for this GC \citep{WC10}.
Spectroscopic studies for NGC~6723, however, are relatively rare. 
\citet{Gra15} have shown the presence of multiple stellar populations in this GC from the Na-O anti-correlation among horizontal branch (HB) stars. 
\citet{Roj16} report chemical abundances for only a few RGB stars.
In order to investigate the chemical properties of RGB stars in NGC~362 and NGC~6723 from our spectroscopy, we have plotted histograms of $\delta$CN, $\delta$HK$'$, and $\delta$CH indices respectively in Figure~\ref{fig_hist}.
These two GCs show clear bimodal distributions in the $\delta$CN index, whereas they show unimodal and narrow distributions in the $\delta$HK$'$ index. 
For the $\delta$CH index, although there are no significant separations, standard deviations of the distributions (0.04 for NGC~362; 0.05 for NGC~6723) are much larger than the typical measurement errors ($\sim$0.017) in both GCs. 
The presence of bimodal CN distribution in NGC~362 is consistent with earlier findings by \citet{Smi84} and \citet{SL09}.
While the CN bimodality in NGC~6723 was not reported previously from spectroscopic observations, \citet{SH86} found some hints of CN variation from DDO photometry.   
\begin{figure*}
\centering
\includegraphics[width=0.85\textwidth]{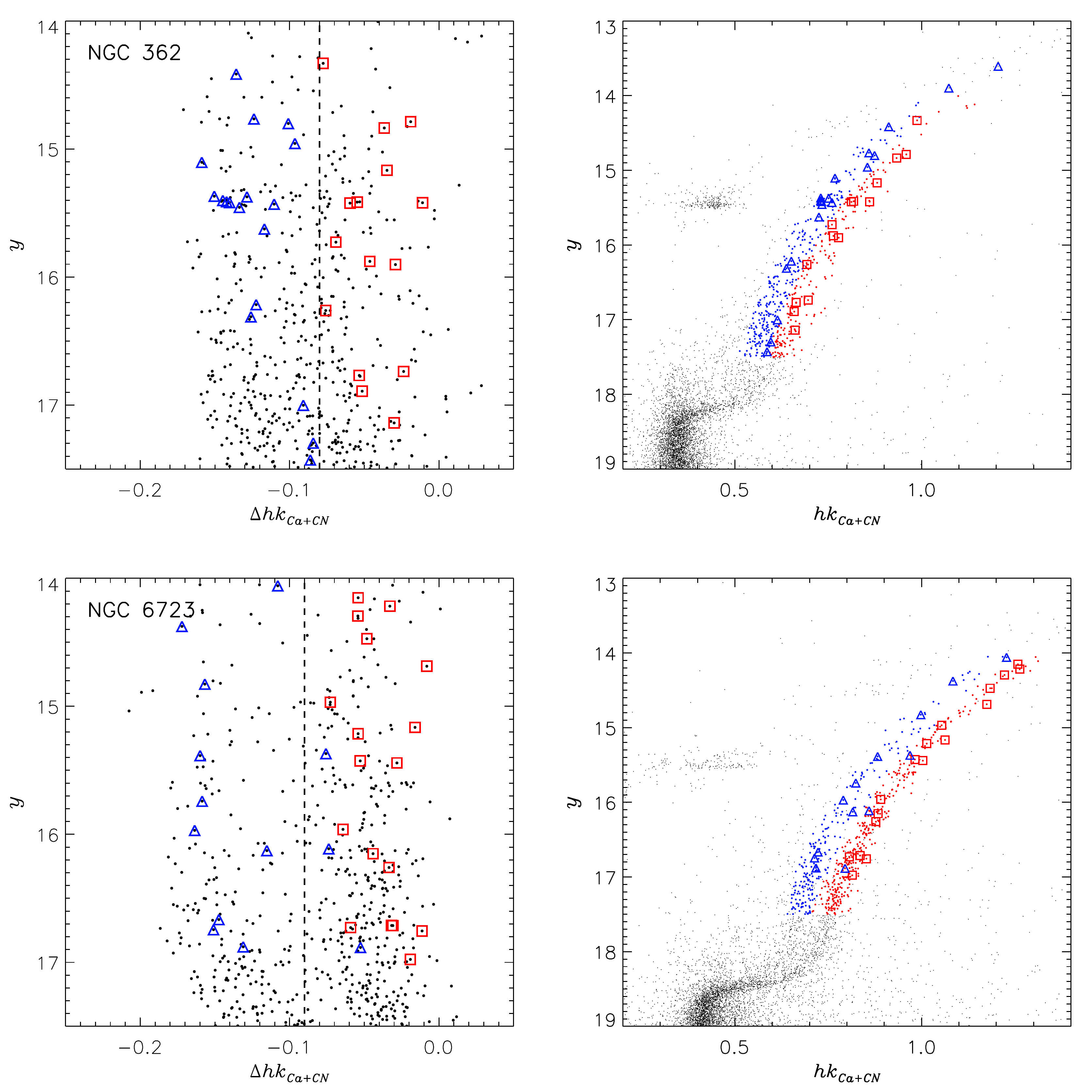}
\figcaption{
Left panels: The RGB stars in NGC~362 (upper) and NGC~6723 (lower) are plotted in the ($y$, $\Delta$$hk_{Ca+CN}$) diagram.
The blue triangles and the red squares represent CN-weak and CN-strong stars, respectively, identified from our low-resolution spectroscopy.
We divided all RGB stars (14.0 $<$ $y$ $<$ 17.5) into two subpopulations, CN-weak and CN-strong, by $\Delta$$hk_{Ca+CN}$=-0.08 (dashed line) for NGC~362 and $\Delta$$hk_{Ca+CN}$=-0.09 for NGC~6723.
Right panels: The two subpopulations are plotted on the ($y$, $hk_{Ca+CN}$) CMD for each GC, where the blue and red points are CN-weak and CN-strong stars, respectively.
\label{fig_subpop}
}
\end{figure*}

For more comprehensive analysis, we have divided observed RGB stars into CN-strong ($\delta$CN $\geq$ 0.0) and CN-weak ($\delta$CN $<$ 0.0) stars for both GCs.
We note that CN-strong and CN-weak stars are clearly separated in our photometry (see Figure~\ref{fig_cmd}), indicating that RGB spread and split in these GCs are due to the difference in CN band strength. 
Figures~\ref{fig_n362spec}--\ref{fig_n6723spec} compare the strengths of measured spectral indices between CN-strong and CN-weak subpopulations for NGC~362 and NGC~6723, respectively.
We note that CN, HK$'$, and CH indices are affected by temperature and gravity, and therefore, we have compared the mean strengths of two subpopulations on the $\delta$-index diagrams (see Section~\ref{obs}).
These two GCs show fairly similar trends in every spectral index.
First of all, as shown in the histogram of Figure~\ref{fig_hist}, CN-strong and CN-weak stars are clearly separated in $\delta$CN index versus $y$ magnitude diagram.
The differences between the two subpopulations are 0.31 for NGC~362, and 0.36 for NGC~6723, which are significant at 14.1$\sigma$ and 10.4$\sigma$ levels respectively, compared to the standard deviation of the mean for each group.
This difference for NGC~362 is comparable to that discovered in previous studies \citep{Smi84,SL09}. 
In case of the $\delta$HK$'$ index, however, the mean values of the two subpopulations are almost identical to within the standard error.
Lastly, the CN-weak stars are more enhanced in the CH band than CN-strong stars, implying the presence of CN-CH anti-correlation, which is a well-known feature in many GCs \citep[e.g.,][]{Har03,Pan10,Smo11}.
For the $\delta$CH index, the two subpopulations are separately by 0.054 (4.4$\sigma$) for NGC~362, and 0.044 (2.5$\sigma$) for NGC~6723.
Our results thus indicate that NGC~362 and NGC~6723 show abundance variations in the light elements (C, N), but not in the heavy element (Ca).
These observations, together with the presence of Na-O anti-correlation and the absence of Fe spread in these GCs \citep{WC10,Car13,Gra15}, suggest that they belong to the ``normal'' GCs with multiple stellar populations.

\begin{figure*}
\centering
\includegraphics[width=0.85\textwidth]{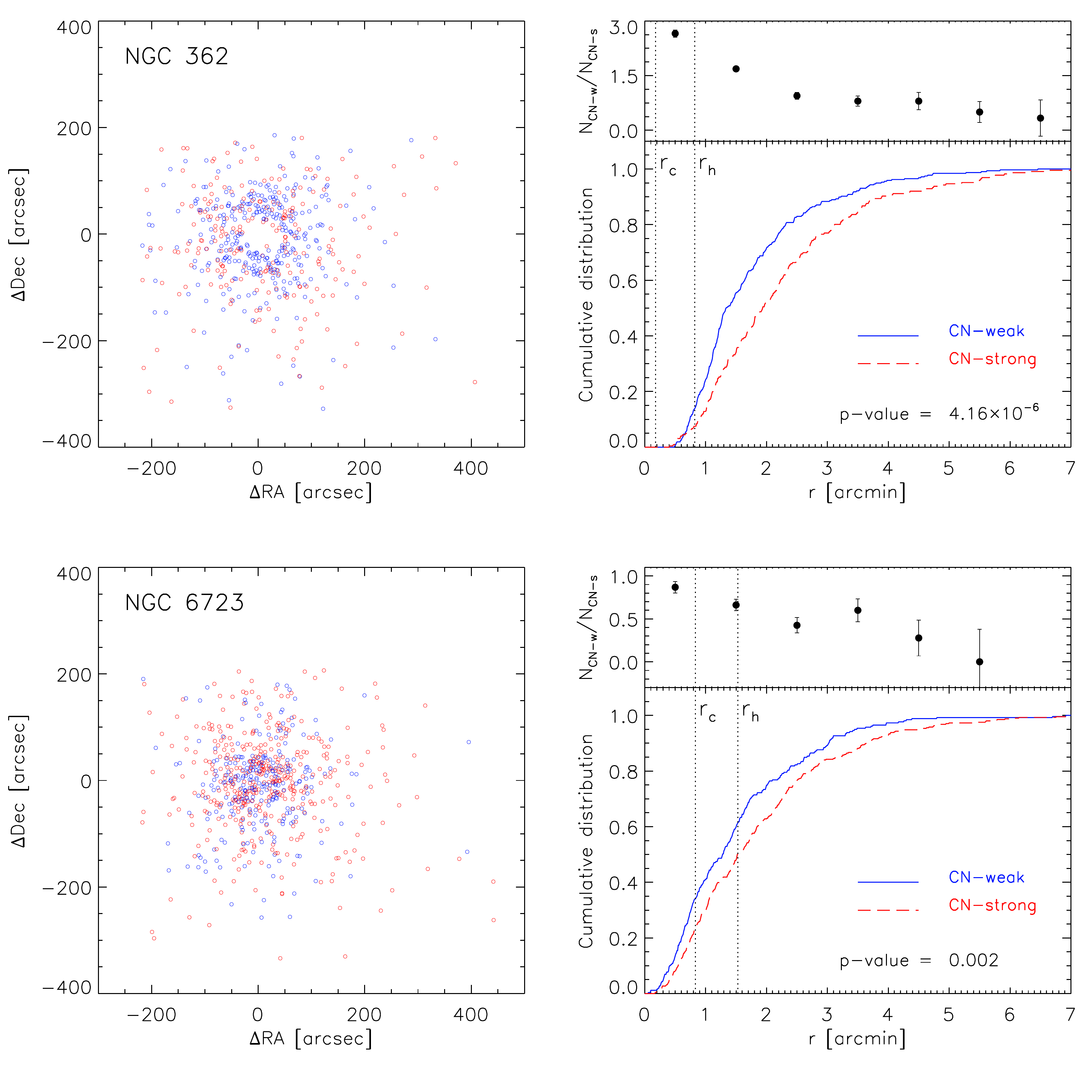}
\figcaption{
Spatial (left panels) and cumulative (right panels) distributions of RGB stars in NGC~362 and NGC~6723, together with number ratio between CN-weak and CN-strong stars as a function of distance from the center.
The blue and red open circles are CN-weak and CN-strong RGB stars identified in the right panels of Figure~\ref{fig_subpop}.
We note that stars placed on the cluster center of NGC~362 ($r$ $<$ 0.5$'$) are excluded by CHI and SHARP parameters (see text).
For both GCs, CN-weak earlier generation stars (blue) are more centrally concentrated than CN-strong later generation stars (red).
The p-value for NGC~362 is only 4.16$\times$10$^{-6}$, indicating that the two subpopulations have definitely different radial distributions. 
In the case of NGC~6723, the p-value is 0.002, which is somewhat larger than that of NGC~362, but still significant.
Vertical dotted lines in right panels mark the core and half-light radii \citep{Har10}.
\label{fig_distri}
}
\end{figure*}
\section{Radial Distributions of Multiple Stellar Populations}\label{rd}
In our low-resolution spectroscopy, observed RGB stars in NGC~362 and NGC~6723 are clearly divided into two subpopulations by CN index, and these subpopulations are also well separated in ($y$, $hk_{Ca+CN}$) CMDs (see Figure~\ref{fig_cmd}). 
However, the number of spectroscopic stars for each GC is limited to only $\sim$30, and we have therefore divided subpopulations using Ca+CN filter photometry in order to secure enough stars to investigate the radial distribution of stellar populations.
First of all, we have plotted ($y$, $\Delta$$hk_{Ca+CN}$) diagrams for the RGB stars including our spectroscopic samples, in the left panels of Figure~\ref{fig_subpop}, where the $\Delta$$hk_{Ca+CN}$ index is determined from the difference between the $hk_{Ca+CN}$ index and the right-edge line of RGB for each GC.
Only RGB stars in the 14.0 $<$ $y$ $<$ 17.5 magnitude range have been used for this analysis.
We have excluded bright stars ($y$ $>$ 14.0) to avoid a possible evolutionary mixing effect, and faint stars ($y$ $<$ 17.5) due to the absence of spectroscopic samples in this magnitude range. 
As shown in these figures, CN-weak (blue triangles) and CN-strong (red squares) stars, classified from our low-resolution spectroscopy, are also well separated in these diagrams.
Therefore, we divided all RGB stars, observed from Ca+CN filter photometry, into CN-weak and CN-strong subpopulations at $\Delta$$hk_{Ca+CN}$=-0.08 for NGC~362, and $\Delta$$hk_{Ca+CN}$=-0.09 for NGC~6723.
These subpopulations are marked on the CMDs in the right panels of Figure~\ref{fig_subpop}.
The number ratio between CN-weak (blue) and CN-strong (red) stars is estimated to be 0.38:0.62 for NGC~6723 (from 597 stars), and 0.59:0.41 for NGC~362 (from 538 stars).
The fraction of enriched subpopulation (CN-strong stars) for NGC~6723 is similar to the general trend of GCs ($\sim$0.68; see \citealt{BL15}), although there is some difference in the definition of enriched subpopulation.

Figure~\ref{fig_distri} shows the spatial distribution (left panels) and the cumulative distribution of each subpopulation (right panels) for RGB stars in NGC~362 and NGC~6723, respectively, together with the number ratio between CN-weak and CN-strong stars as a function of distance from the center.
The stars placed on the cluster center of NGC~362 ($r$ $<$ 0.5$'$) are excluded by CHI and SHARP parameters due to the severe contamination by adjacent starlight (see Section~\ref{obs}).
In this figure, the CN-weak stars (blue) are more centrally concentrated than the CN-strong stars (red) in NGC~362, and the number ratio decreases with increasing distance from the center. 
In general, CN-weak stars are considered to be earlier generation stars, and therefore, this result is contrary to the fact that many GCs show a central concentration of later generation stars (see Section~\ref{intro}).
In order to verify the difference in radial distribution between the two subpopulations, we have performed the Kolmogorov-Smirnov (KS) test.
The probability-value (p-value) is only 4.16$\times$10$^{-6}$ with a maximum deviation of 0.22 for NGC~362.
This result thus indicates that these two subpopulations definitely have different radial distributions.
NGC~6723 also shows a central concentration of CN-weak earlier generation stars.
The KS test indicates the p-value of 0.002 and the deviation of 0.15, confirming that the two subpopulations show different radial distributions.
When we divide RGB stars into three subpopulations, CN-weak, CN-intermediate, and CN-strong, the central concentration of CN-weak subpopulation becomes more clear in both GCs. 
\begin{figure}
\centering
\includegraphics[width=0.45\textwidth]{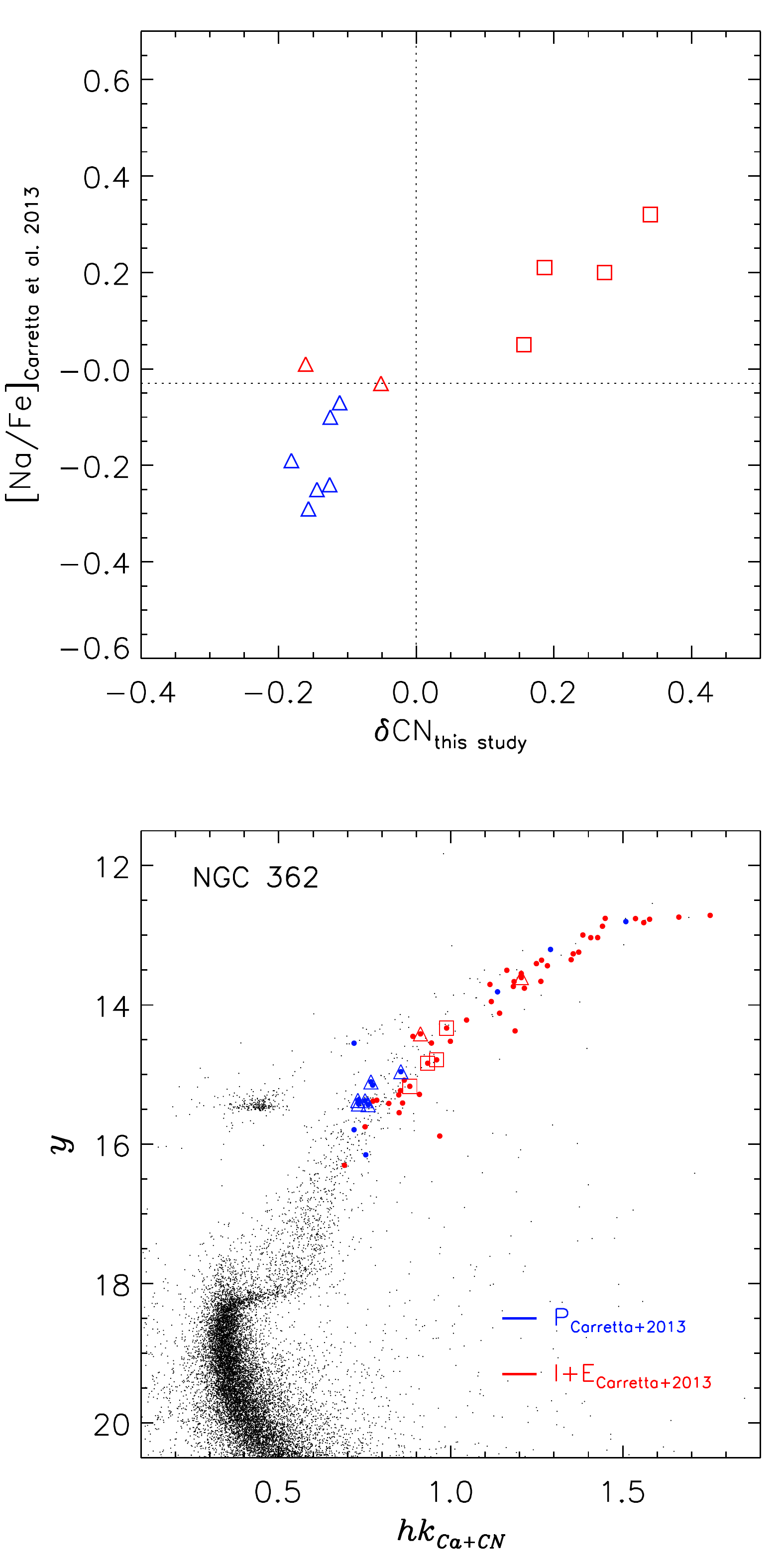}
\figcaption{
Comparison between our study and C13 for NGC~362. 
Upper panel: The [Na/Fe] abundance from C13 plotted against the $\delta$CN index of this study for 12 common stars. 
We note that they show a strong correlation and similar subgrouping.
The blue and red colors represent the first (primordial) and second (intermediate and extreme) generation stars, divided by [Na/Fe], and triangles and squares indicate the CN-weak and CN-strong stars, divided by $\delta$CN index.
Dotted lines denote the criteria of each study ($\delta$CN=0.0; [Na/Fe]=-0.03).
Lower panel: C13 stars are identified on the ($y$, $hk_{Ca+CN}$) CMD, where the blue and red circles are the first and second generation stars, respectively.
It is important to note that most of the bright stars ($y$ $<$ 14.0) in C13 are defined as the second generation, while these stars are excluded in our study (see text). 
\label{fig_carretta}
}
\end{figure}

\citet[][hereafter C13]{Car13}, however, report an apparently different result for the radial distribution of stars in NGC~362.
In their high-resolution spectroscopy, the second generation stars (intermediate and extreme components) show a more centrally concentrated distribution than the first generation stars (primordial component), where primordial, intermediate, and extreme components are defined from the Na-O anti-correlation.
As described by them, however, the level of confidence for the radial distribution is not very high due to the small sample size (N=71).
The p-value\footnote{``KS-probability'' in C13.} is 0.2730, which is much larger than the value from our result (4.16$\times$10$^{-6}$).
On the other hand, it is important to check the definition of subpopulations in the two studies, because we have divided the two subpopulations by CN index, whereas C13 was based on [Na/Fe] abundance.
In the upper panel of Figure~\ref{fig_carretta}, we have compared our $\delta$CN index with [Na/Fe] abundance of C13 for 12 common spectroscopic stars.
They show a strong correlation, which is in good agreement with previous studies for other GCs \citep{Sne92,Mar08}, indicating that the subgrouping of our study is not significantly different from that of C13.

The number ratio between the earlier and later generation stars obtained from this study for NGC~362 is also different from that suggested by C13.
The fraction of CN-weak earlier generation stars estimated from our Ca+CN photometry is $\sim$0.59, while that of primordial component stars from C13 is only $\sim$0.22.
In the lower panel of Figure~\ref{fig_carretta}, we have cross-matched stars of C13 on the ($y$, $hk_{Ca+CN}$) CMD to find the origin of this difference. 
As shown in this CMD, most of the bright stars ($y$ $<$ 14.0) in C13 are classified as the later generation, but our study excluded these stars from the analysis to avoid a possible evolutionary mixing effect.
Therefore, the different results between the two studies might be due to the classification of these bright stars.  
However, for more rigorous comparison, it is required to investigate the correlation between stellar populations divided by CN index and those divided in the Na-O plane.

\section{Discussion}\label{dis}
We have shown that the RGB stars in NGC~362 and NGC~6723 are clearly divided into two subpopulations by CN index, but these two subpopulations show no difference in calcium abundance from our low-resolution spectroscopy.
The well-known CN-CH anti-correlation between the two subpopulations is also shown in both GCs.
These results, together with previous findings by other investigators \citep{WC10,Car13,Gra15}, suggest that these two GCs are ``normal'' GCs with multiple stellar populations.
Furthermore, we found that the CN-weak earlier generation stars are significantly more centrally concentrated than CN-strong later generation stars in both GCs.
These findings are important as the second and third cases that show such a reversed radial distribution trend, following M15 by \citet{Lar15}.
We note, however, that the innermost region ($r$ $<$ 0.5$'$) of a cluster was not included in our analysis for NGC~362, and therefore further analysis with the recent HST UV survey \citep{Pio15} would be helpful to confirm the reversed spatial distribution discovered in this study.

As described in Section~\ref{intro}, the self-enrichment scenario predicts more centrally concentrated stars belonging to the later generation \citep[see, e.g.,][]{Ves13}. 
Therefore, our finding of centrally concentrated CN-weak earlier generation stars in NGC~362 and NGC~6723 is not generally acceptable in this framework. 
The merging of two individual GCs might be a possible solution for different radial distributions of stellar populations \citep[see][]{Car11}, but these GCs do not show additional evidence of merging, such as metallicity difference. 
Moreover, their chemical properties, the bimodal CN distribution and CN-CH anti-correlation without Fe spread, are not intuitively understandable in the merging scenario. 
\citet{Lar15} suggested that mass segregation may explain the observed radial trend of M15.
This scenario, however, requires extreme He enhancement of later generation stars ($Y$ $\geq$ 0.40) in order to produce a mass difference of 0.25$M_{\odot}$, which is not observed in this GC. 
In either case of NGC~362 and NGC~6723, He enhancement is not yet reported, although some $\Delta$$Y$ is expected. 
Therefore, the mass segregation by the difference in He abundance alone is not likely to explain the observed radial trends in M15, NGC~362, and NGC~6723. 
More recently, an alternative solution for the mass difference is proposed by \citet{Hen15}, which suggests that the different incidence of binary from primordial and enriched stellar populations may provoke this required mass difference \citep[see also][]{Hong15}. 
According to them, this effect is more efficient in more concentrated GCs. 
Interestingly, M15 has been known as a core collapsed GC, and NGC~362 and NGC~6723 are also classified as possible core collapsed GCs \citep{Har10}. 
The effects of mass segregation on the radial distribution of stellar generations require more detailed studies. 
To this end, we plan to run direct $N$-body simulations of NGC~362 and NGC~6723 (Pasquato et al. in prep.) under different assumptions for the
relevant binary fractions, initial concentration of the second generation stars, and mass-difference between the subpopulations to assess to what extent mass-segregation can explain our findings in these GCs. 
At this stage, it is worth noting that the half-mass relaxation time of NGC~362 is relatively short (less than 1Gyr) while that of NGC~6723 is roughly of the same order of that of M15 ($\sim$2Gyr). 
Therefore, we expect that mass-segregation would play a more important role in the former cluster.

One caveat in this analysis is that definitions of subpopulations are not exactly identical in many studies.
We have divided stars in a GC into two subpopulations by the strength of the CN band, whereas many high-resolution spectroscopic studies based on Na and O abundances prefer to separate them into three subpopulations \citep{Car09,JP12}.
In addition, \citet{Jang14} and \citet{JL15} suggest the presence of three subpopulations with different He abundances based on population models for RR Lyrae and HB stars.
As shown in our previous study \citep{Lim15}, CN-weak stars in NGC~1851 could be further divided into two subpopulations, and therefore, it is possible that CN-weak stars in other GCs also might be further divided into two subgroups.
This issue will be discussed in our forthcoming paper. 

\acknowledgments 
We are grateful to the anonymous referee for a number of helpful suggestions.
Support for this work was provided by the National Research Foundation of Korea to the Center for Galaxy Evolution Research, and by KASI under the R\&D program (Project No. 2014-1-600-05) supervised by the Ministry of Science, ICT and future Planning.
M.P. acknowledges support from Mid-career Researcher Program (No. 2015-008049) through the National Research Foundation (NRF) of Korea.


\begin{deluxetable*}{cccccccccccccc}
\setlength{\tabcolsep}{0.04in}
\tabletypesize{\scriptsize}
\tablewidth{0pt}
\tablecaption{Index Measurements for the sample stars in NGC~362 and NGC~6723\label{tab_index}}
\tablehead{
\colhead{ID} & \colhead{Ra} & \colhead{Dec} & \colhead{$y$} & \colhead{$hk_{Ca+CN}$} & \colhead{HK$'$} & \colhead{errHK$'$} & \colhead{$\delta$HK$'$} & \colhead{CN} & \colhead{errCN} & \colhead{$\delta$CN} & \colhead{CH} & \colhead{errCH} & \colhead{$\delta$CH}
}
\startdata
N362-1003 &  15.82759 & -70.89803 & 13.8980 &  1.0730 &  0.6215 &  0.0147 &  0.0304 & -0.0471 &  0.0228 & -0.0379 &  1.2000 &  0.0104 & -0.0093 \\
N362-1005 &  15.70891 & -70.86282 & 14.4150 &  0.9120 &  0.4938 &  0.0170 & -0.0648 & -0.2056 &  0.0272 & -0.1609 &  1.1570 &  0.0113 & -0.0239 \\
N362-1009 &  15.87088 & -70.89255 & 14.7650 &  0.8590 &  0.6008 &  0.0196 &  0.0643 & -0.0910 &  0.0311 & -0.0222 &  1.2243 &  0.0135 &  0.0626 \\
N362-1014 &  15.91741 & -70.82832 & 15.1040 &  0.7680 &  0.5073 &  0.0221 & -0.0078 & -0.2171 &  0.0360 & -0.1250 &  1.1537 &  0.0149 &  0.0106 \\
N362-1019 &  15.92921 & -70.91448 & 15.3690 &  0.7300 &  0.4985 &  0.0368 &  0.0001 & -0.2362 &  0.0606 & -0.1260 &  1.1697 &  0.0245 &  0.0412 \\
N362-1020 &  16.02489 & -70.87573 & 15.3740 &  0.7510 &  0.4916 &  0.0225 & -0.0065 & -0.2922 &  0.0384 & -0.1816 &  1.1890 &  0.0147 &  0.0608 \\
N362-1022 &  15.79167 & -70.90933 & 15.4020 &  0.7300 &  0.5133 &  0.0293 &  0.0170 & -0.2802 &  0.0503 & -0.1677 &  1.1753 &  0.0196 &  0.0486 \\
N362-1023 &  15.84620 & -70.92680 & 15.4120 &  0.7310 &  0.4520 &  0.0376 & -0.0436 & -0.2891 &  0.0623 & -0.1758 &  1.1804 &  0.0241 &  0.0542 \\
N362-1024 &  15.72000 & -70.86794 & 15.4190 &  0.7320 &  0.4988 &  0.0274 &  0.0036 & -0.2705 &  0.0462 & -0.1568 &  1.1654 &  0.0183 &  0.0396 \\
N362-1028 &  15.94930 & -70.83397 & 15.4560 &  0.7330 &  0.4782 &  0.0239 & -0.0147 & -0.2764 &  0.0400 & -0.1601 &  1.1615 &  0.0158 &  0.0378 \\
N362-1529 &  15.81924 & -70.82098 & 14.3310 &  0.9880 &  0.5428 &  0.0189 & -0.0211 &  0.1474 &  0.0246 &  0.1863 &  1.1771 &  0.0129 & -0.0084 \\
N362-2003 &  16.00982 & -70.81965 & 13.6070 &  1.2050 &  0.6492 &  0.0110 &  0.0397 & -0.0406 &  0.0174 & -0.0514 &  1.1866 &  0.0080 & -0.0386 \\
N362-2011 &  15.63266 & -70.82576 & 14.7870 &  0.9590 &  0.5600 &  0.0207 &  0.0249 &  0.2696 &  0.0251 &  0.3399 &  1.1316 &  0.0147 & -0.0289 \\
N362-2012 &  15.96672 & -70.88182 & 14.8010 &  0.8740 &  0.5360 &  0.0182 &  0.0018 & -0.1339 &  0.0285 & -0.0626 &  1.1684 &  0.0124 &  0.0087 \\
N362-2014 &  16.04104 & -70.82831 & 14.8360 &  0.9330 &  0.5448 &  0.0152 &  0.0128 &  0.2000 &  0.0190 &  0.2736 &  1.0482 &  0.0111 & -0.1096 \\
N362-2017 &  15.64483 & -70.88963 & 14.9570 &  0.8550 &  0.5397 &  0.0203 &  0.0153 & -0.1934 &  0.0332 & -0.1114 &  1.1392 &  0.0141 & -0.0120 \\
N362-2025 &  15.67090 & -70.93616 & 15.1670 &  0.8810 &  0.4925 &  0.0308 & -0.0186 &  0.0599 &  0.0411 &  0.1563 &  1.1330 &  0.0208 & -0.0066 \\
N362-2038 &  15.87910 & -70.90480 & 15.4150 &  0.8180 &  0.5039 &  0.0263 &  0.0085 &  0.1143 &  0.0340 &  0.2277 &  1.0903 &  0.0184 & -0.0357 \\
N362-2039 &  16.07741 & -70.87669 & 15.4200 &  0.8610 &  0.4768 &  0.0205 & -0.0183 &  0.0389 &  0.0274 &  0.1527 &  1.0962 &  0.0140 & -0.0295 \\
N362-2040 &  15.97682 & -70.83964 & 15.4230 &  0.8120 &  0.4920 &  0.0224 & -0.0030 & -0.0000 &  0.0311 &  0.1139 &  1.1459 &  0.0150 &  0.0203 \\
N362-2041 &  16.05219 & -70.79980 & 15.4310 &  0.7600 &  0.4845 &  0.0199 & -0.0100 & -0.2590 &  0.0330 & -0.1445 &  1.1472 &  0.0133 &  0.0221 \\
N362-3006 &  15.63587 & -70.83836 & 16.2160 &  0.6510 &  0.4362 &  0.0273 & -0.0088 & -0.3131 &  0.0457 & -0.1446 &  1.1133 &  0.0181 &  0.0312 \\
N362-3007 &  15.91064 & -70.82910 & 16.3110 &  0.6380 &  0.4114 &  0.0276 & -0.0276 & -0.3737 &  0.0474 & -0.1987 &  1.0984 &  0.0181 &  0.0216 \\
N362-4004 &  15.66253 & -70.88970 & 15.6250 &  0.7250 &  0.4703 &  0.0197 & -0.0119 & -0.3182 &  0.0338 & -0.1903 &  1.1331 &  0.0132 &  0.0186 \\
N362-4009 &  15.76517 & -70.93929 & 15.7280 &  0.7600 &  0.4932 &  0.0463 &  0.0174 & -0.0302 &  0.0657 &  0.1047 &  1.1531 &  0.0310 &  0.0443 \\
N362-4015 &  15.67688 & -70.85783 & 15.8780 &  0.7640 &  0.4750 &  0.0218 &  0.0087 & -0.0052 &  0.0300 &  0.1401 &  1.0739 &  0.0151 & -0.0266 \\
N362-4017 &  15.94820 & -70.91762 & 15.9030 &  0.7780 &  0.4590 &  0.0275 & -0.0057 &  0.0227 &  0.0368 &  0.1697 &  1.0834 &  0.0188 & -0.0158 \\
N362-4033 &  15.87707 & -70.89821 & 16.2620 &  0.6930 &  0.3905 &  0.0325 & -0.0515 & -0.0603 &  0.0441 &  0.1113 &  1.0471 &  0.0216 & -0.0323 \\
N362-4060 &  16.06938 & -70.80835 & 16.7380 &  0.6960 &  0.3462 &  0.0492 & -0.0659 & -0.0844 &  0.0658 &  0.1199 &  0.9544 &  0.0337 & -0.0990 \\
N362-4062 &  15.98650 & -70.84545 & 16.7690 &  0.6640 &  0.3982 &  0.0395 & -0.0119 &  0.0026 &  0.0516 &  0.2091 &  0.9687 &  0.0278 & -0.0830 \\
N362-4072 &  15.69080 & -70.80397 & 16.8920 &  0.6590 &  0.4152 &  0.0424 &  0.0129 &  0.0135 &  0.0556 &  0.2285 &  1.0207 &  0.0292 & -0.0242 \\
N362-4088 &  16.02500 & -70.90798 & 17.0020 &  0.6140 &  0.4306 &  0.0505 &  0.0352 & -0.3890 &  0.0893 & -0.1665 &  1.0825 &  0.0338 &  0.0436 \\
N362-4109 &  16.03894 & -70.90070 & 17.1400 &  0.6610 &  0.4006 &  0.0495 &  0.0139 & -0.1198 &  0.0704 &  0.1122 &  1.0017 &  0.0341 & -0.0295 \\
N362-4138 &  16.01084 & -70.85294 & 17.2970 &  0.5960 &  0.3716 &  0.0595 & -0.0052 & -0.4562 &  0.1060 & -0.2135 &  1.0409 &  0.0393 &  0.0182 \\
N362-4162 &  15.64587 & -70.87316 & 17.4290 &  0.5860 &  0.4590 &  0.0477 &  0.0905 & -0.3001 &  0.0801 & -0.0482 &  1.0445 &  0.0332 &  0.0291 \\
N6723-1004 & 284.85938 & -36.62233 & 14.3757 &  1.0840 &  0.6275 &  0.0167 &  0.0031 & -0.0150 &  0.0256 & -0.2439 &  1.1656 &  0.0121 &  0.0362 \\
N6723-1007 & 284.86337 & -36.67758 & 15.3852 &  0.8823 &  0.5005 &  0.0257 & -0.0496 & -0.1947 &  0.0410 & -0.2946 &  1.1313 &  0.0175 &  0.0242 \\
N6723-1026 & 284.91367 & -36.67157 & 14.0584 &  1.2275 &  0.6672 &  0.0170 &  0.0195 & -0.0693 &  0.0276 & -0.3386 &  1.1832 &  0.0125 &  0.0467 \\
N6723-1029 & 284.92978 & -36.65347 & 14.8267 &  0.9979 &  0.5989 &  0.0198 &  0.0077 & -0.1008 &  0.0316 & -0.2720 &  1.1651 &  0.0141 &  0.0456 \\
N6723-2001 & 284.81909 & -36.71124 & 14.2177 &  1.2627 &  0.6510 &  0.0150 &  0.0150 &  0.4101 &  0.0175 &  0.1611 &  1.0783 &  0.0116 & -0.0546 \\
N6723-2002 & 284.83209 & -36.61426 & 15.2139 &  1.0138 &  0.5649 &  0.0208 &  0.0022 &  0.2973 &  0.0247 &  0.1756 &  1.1007 &  0.0150 & -0.0102 \\
N6723-2017 & 284.87149 & -36.58863 & 14.9679 &  1.0537 &  0.5728 &  0.0172 & -0.0080 &  0.2545 &  0.0212 &  0.1013 &  1.1359 &  0.0122 &  0.0195 \\
N6723-2025 & 284.88257 & -36.64808 & 14.2945 &  1.2221 &  0.6571 &  0.0196 &  0.0268 &  0.3931 &  0.0232 &  0.1539 &  1.0984 &  0.0150 & -0.0328 \\
N6723-2037 & 284.90323 & -36.58818 & 14.1520 &  1.2579 &  0.6734 &  0.0144 &  0.0326 &  0.4082 &  0.0171 &  0.1508 &  1.0640 &  0.0114 & -0.0704 \\
N6723-2044 & 284.91840 & -36.57550 & 14.4722 &  1.1837 &  0.6322 &  0.0145 &  0.0149 &  0.3329 &  0.0176 &  0.1164 &  1.0888 &  0.0110 & -0.0385 \\
N6723-2048 & 284.93604 & -36.68473 & 14.6869 &  1.1745 &  0.6238 &  0.0160 &  0.0223 &  0.4239 &  0.0181 &  0.2349 &  1.1039 &  0.0119 & -0.0186 \\
N6723-2050 & 284.94626 & -36.60479 & 15.4423 &  1.0034 &  0.5273 &  0.0230 & -0.0186 &  0.2299 &  0.0279 &  0.1373 &  1.0220 &  0.0170 & -0.0839 \\
N6723-2051 & 284.96124 & -36.63554 & 15.4265 &  0.9820 &  0.5186 &  0.0270 & -0.0285 &  0.0999 &  0.0356 &  0.0054 &  1.1630 &  0.0183 &  0.0567 \\
N6723-2052 & 284.96408 & -36.58178 & 15.3696 &  0.9689 &  0.5155 &  0.0234 & -0.0357 &  0.0985 &  0.0310 & -0.0033 &  1.1302 &  0.0161 &  0.0227 \\
N6723-2054 & 285.01886 & -36.67313 & 15.1647 &  1.0628 &  0.4544 &  0.0207 & -0.1119 &  0.1629 &  0.0251 &  0.0349 &  1.1812 &  0.0133 &  0.0692 \\
N6723-3001 & 284.81351 & -36.57934 & 16.1273 &  0.8157 &  0.5186 &  0.0235 &  0.0231 & -0.0885 &  0.0352 & -0.0935 &  1.1441 &  0.0161 &  0.0534 \\
N6723-3003 & 284.82947 & -36.66508 & 16.8789 &  0.7171 &  0.4734 &  0.0337 &  0.0331 & -0.3082 &  0.0577 & -0.2172 &  1.1146 &  0.0228 &  0.0405 \\
N6723-3046 & 284.88876 & -36.59848 & 15.7411 &  0.8238 &  0.5142 &  0.0232 & -0.0098 & -0.2194 &  0.0381 & -0.2738 &  1.1424 &  0.0158 &  0.0432 \\
N6723-3063 & 284.90668 & -36.58289 & 15.9691 &  0.7903 &  0.5167 &  0.0217 &  0.0095 & -0.2719 &  0.0371 & -0.2971 &  1.1454 &  0.0148 &  0.0512 \\
N6723-3073 & 284.91208 & -36.58960 & 16.7448 &  0.7132 &  0.4585 &  0.0301 &  0.0084 & -0.3149 &  0.0511 & -0.2411 &  1.0876 &  0.0204 &  0.0105 \\
N6723-3091 & 284.95197 & -36.62900 & 16.6659 &  0.7227 &  0.4666 &  0.0334 &  0.0107 & -0.2883 &  0.0562 & -0.2245 &  1.1214 &  0.0224 &  0.0425 \\
N6723-4045 & 284.86853 & -36.60382 & 16.7289 &  0.8062 &  0.4552 &  0.0349 &  0.0039 &  0.1034 &  0.0441 &  0.1752 &  1.0160 &  0.0247 & -0.0615 \\
N6723-4077 & 284.89359 & -36.60166 & 15.9613 &  0.8906 &  0.5287 &  0.0214 &  0.0210 &  0.2933 &  0.0250 &  0.2671 &  1.1161 &  0.0150 &  0.0217 \\
N6723-4121 & 284.91809 & -36.64885 & 16.1128 &  0.8593 &  0.4993 &  0.0197 &  0.0027 & -0.0401 &  0.0282 & -0.0470 &  0.9698 &  0.0149 & -0.1213 \\
N6723-4124 & 284.92496 & -36.64448 & 16.7130 &  0.8357 &  0.4640 &  0.0335 &  0.0115 &  0.1229 &  0.0421 &  0.1927 &  1.0108 &  0.0239 & -0.0670 \\
N6723-4128 & 284.92929 & -36.61201 & 16.1512 &  0.8832 &  0.5046 &  0.0237 &  0.0108 &  0.1575 &  0.0297 &  0.1555 &  1.1295 &  0.0162 &  0.0393 \\
N6723-4131 & 284.93518 & -36.64169 & 16.2592 &  0.8777 &  0.4721 &  0.0250 & -0.0137 &  0.1503 &  0.0309 &  0.1621 &  1.0928 &  0.0171 &  0.0049 \\
N6723-4136 & 284.94473 & -36.72401 & 16.9771 &  0.8143 &  0.4696 &  0.0471 &  0.0366 &  0.0509 &  0.0625 &  0.1544 &  1.0649 &  0.0327 & -0.0070 \\
N6723-4148 & 284.96790 & -36.70012 & 16.7123 &  0.8349 &  0.4840 &  0.0320 &  0.0315 &  0.2367 &  0.0377 &  0.3064 &  0.9647 &  0.0238 & -0.1131 \\
N6723-4153 & 284.97528 & -36.68954 & 16.8838 &  0.7949 &  0.4130 &  0.0345 & -0.0269 & -0.2320 &  0.0536 & -0.1404 &  1.1074 &  0.0225 &  0.0334 \\
N6723-4162 & 285.04105 & -36.68499 & 16.7548 &  0.8521 &  0.4049 &  0.0371 & -0.0445 & -0.0728 &  0.0513 &  0.0023 &  1.0944 &  0.0243 &  0.0175
\enddata
\end{deluxetable*}

\end{document}